\documentclass[draft]{journal_style}
\usepackage{url} %this package should fix any errors with URLs in refs.
\usepackage{soul}
 \usepackage{float}
 \usepackage{subcaption}
 \usepackage{multirow}
\draftfalse

\begin{document}
\title{Unstable drainage dynamics during multiphase flow across capillary heterogeneities}

\authors{C. Harris\affil{1}, S. Krevor\affil{1}, A. H. Muggeridge\affil{1}, M. Camilleri\affil{2} and S. J. Jackson\affil{2}}

\affiliation{1}{Department of Earth Science and Engineering, Imperial College London, London, UK}
\affiliation{2}{CSIRO Energy, Clayton North, Victoria, Australia}

\correspondingauthor{S. J. Jackson}{samuel.jackson@csiro.au}

\begin{keypoints}
\item Real time imaging captures multiscale fluid dynamics in heterogeneous porous materials, enabling new studies of complex media
\item Direct observations of multiphase flow through layered rock shows subtle connectivity differences drive unstable flow and variable breakthrough times
\item Amplified uncertainty at large scales challenges deterministic models and supports probabilistic frameworks to capture natural variability
\end{keypoints}

\section*{\centering Abstract}
We use novel, fast 4D Synchrotron X-ray imaging with large field-of-view to reveal pore- and macro-scale drainage dynamics during gas–brine flow through a layered sandstone rock sample. We show that a single centimetre-scale layer, similar in pore size distribution to the surrounding rock but with reduced connectivity, temporarily inhibits and redirects gas flow, acting as a capillary barrier. Subtle variations in gas invasion upstream of the barrier lead to different downstream migration pathways over repeated experiments, resulting in unstable and unpredictable drainage behaviour, with breakthrough times varying by up to a factor of four. The results show that heterogeneity in pore-scale connectivity can amplify variability in macroscopic flow, challenging deterministic assumptions in existing continuum models. By linking structural heterogeneity to flow instability, this work underscores the need for probabilistic modelling approaches in multiphase flow and highlights broader implications for managing fluid transport in natural and engineered porous systems.

\section*{\centering Plain Language Summary}
We use a novel real-time imaging technique to capture fluid dynamics across multiple scales within opaque, heterogeneous porous materials. By directly observing gas–brine flow through a layered centimetre-scale rock sample, it reveals how subtle differences in microscopic connectivity can drive macroscopic flow instability and amplify uncertainty at larger scales. These insights challenge deterministic modelling of continuum multiphase flow in porous media and underscore the need for probabilistic frameworks that better account for natural variability. Additionally, the demonstrated imaging techniques offer new opportunities for investigating multiscale dynamics in complex media, with broad applications across geoscience, energy systems, and environmental remediation.

\section{Introduction}
Multiphase flow through heterogeneous porous materials underpins critical processes in environmental science, energy systems, and subsurface engineering, including groundwater flow, energy storage and geological carbon storage. Accurate modeling of fluid flow is of particular importance for the geological storage of carbon dioxide (CO$_2$), whereby conformance of the plume migration with forecasts from numerical models is essential for environmental assurance and regulatory compliance \cite{chadwick2015}. Yet, at many storage sites CO$_2$ migration has proved difficult to predict \cite{Jackson2020}, largely due to the multiscale heterogeneity of the rock. Beyond large-scale variations in porosity and permeability, subtle differences in pore connectivity give rise to capillary pressure variations that can strongly influence plume migration and trapping \cite{Regnier2019, Kortekaas1985, Ringrose1993, Li2015,Jackson2018,Krevor2011}. These capillary heterogeneities can act as local barriers, redirecting, delaying, or immobilising CO$_2$ \cite{Ni2023, Bump2023, Jackson2020, Trevisan2017a, Saadatpoor2010,Krevor2011,Bech2018,Harris2021}, leading to baffled migration paths and locally elevated saturations observed at field sites such as Sleipner \cite{Chadwick2009,Fernø2023,Ni2021,Jackson2020REV,Seyyedi2022}. 

Recent experiments further show that these processes are both spatially variable and dynamically complex \cite{Jackson2018, Krevor2011, Spurin2020, Fernø2023}. Fluid flow through heterogeneous porous media may be unsteady and non-unique, highlighting the need to identify the underlying mechanisms that govern flow variability \cite{Chadwick2009}. Despite advances in continuum-scale modelling, key questions remain about how pore-scale displacement mechanisms give rise to macroscopic flow behaviour. While the dynamics within individual pore throats are critical \cite{Rucker2015, Spurin2020}, it is their collective behaviour across thousands of interconnected pores that determines large-scale outcomes \cite{Jackson2020}. Traditional imaging techniques lack the spatial or temporal resolution to resolve these multiscale processes in representative rock samples. As a result, most prior work has focused on simplified porous media systems such as synthetic micro-models \cite{maloy1992, Zhao2016, Moura2020} or focused solely on pore-scale dynamics \cite{Singh2017, Spurin2020} and steady-state regimes \cite{Bultreys2018, Jackson2020REV}. Bridging this gap requires new experimental approaches capable of linking pore-scale dynamics with macroscopic flow. 

In this work, we use high-speed 4D Synchrotron X-ray micro-computed tomography at the Australian Synchrotron to directly image gas–brine displacement in a sandstone with natural layered heterogeneity. Our set-up achieves micrometre-scale spatial resolution across centimetre-scale samples, with time resolution on the order of minutes, simultaneously capturing dynamic pore-scale events and their influence on macroscopic flow behaviour. This capability is enabled by recent advances in detector field-of-view, synchrotron photon flux, and big-data image processing. We show that subtle differences in upstream invasion can lead to markedly different flow paths and breakthrough times. These results provide direct evidence of unstable, non-unique behaviour emerging from fine-scale capillary heterogeneity under nominally identical, macroscopic boundary conditions. Our findings demonstrate that small-scale capillary heterogeneity can amplify large-scale flow uncertainty, supporting the case for stochastic approaches in multiphase flow modelling. More broadly, the imaging approach introduced here provides a powerful tool for investigating transient multiscale flow processes in complex, opaque porous systems, relevent across geoscience, energy, and environmental applications.

\section{Materials \& Methods}

An experiment was designed to capture micron resolution dynamics within a heterogeneous sample over a centimeter-scale field-of-view. Vertical coreflood drainage experiments were repeated at two distinct flow rates on a heterogeneous Bentheimer sandstone rock sample containing low porosity layers oblique to flow. Its dimensions (diameter $\sim$ 12.35 mm, length $\sim$ 64.7 mm, volume $\sim$ 7750 mm\textsuperscript{3}) were chosen to compliment the imaging setup, which was optimised to capture pore-scale flow resolved over many representative elementary volumes (IMBL field of view $\sim$ 6 cm, spatial resolution $\sim$ 20 $\mu$m,  temporal resolution $\sim$ 7 minutes). The sample had previously been characterised extensively through steady-state experiments and modelling \cite{Jackson2020REV, Zahasky2020a, Jackson2022}. The shallow-marine sandstone consists of $\sim$95\% quartz with minor feldspar and clay \cite{Peksa2015}. Core-averaged permeability is 0.681 $\pm$ 0.006 D \cite{Jackson2020REV} and porosity is 0.19 $\pm$ 0.01, with 99\% of the porosity connected, as determined from the segmentation workflow (see Supporting Information).

Supercritical nitrogen (N$_2$) and brine (3.5 wt.\% KI) were used as non-wetting and wetting phases respectively. Nitrogen was used as an analogue to supercritical CO\textsubscript{2}, owing to its comparable immiscible fluid properties relevant to subsurface carbon storage \cite{Krevor2011, Niu2015}. The KI brine provided sufficient absorption contrast between fluids while also enhancing phase contrast of the grain structure.

\begin{table}[h]
\caption{Experimental flow rate and associated pore-scale capillary number. The duration of fluid injection is displayed in minutes and pore volumes injected (PVI), with the number of scans taken listed}
     \label{tab:SyncExp_timings} 
\centering
\begin{tabular}{|c|c|c|c|ccc|}
\hline
\multirow{2}{*}{\begin{tabular}[c]{@{}c@{}}Experiment\\ name\end{tabular}} & \multirow{2}{*}{\begin{tabular}[c]{@{}c@{}}Rate \\ {[}ml/min{]}\end{tabular}} & \multirow{2}{*}{\begin{tabular}[c]{@{}c@{}}Capillary \\ number {[}-{]}\end{tabular}} & \multirow{2}{*}{\begin{tabular}[c]{@{}c@{}}Repeat \\ name\end{tabular}} & \multicolumn{3}{c|}{Injection duration} \\ \cline{5-7} 
 &  &  &  & \multicolumn{1}{c|}{{[}min{]}} & \multicolumn{1}{c|}{{[}PVI{]}} & {[}\# scans{]} \\ \hline
\multirow{2}{*}{HF} & \multirow{2}{*}{0.1} & \multirow{2}{*}{4.9x10$^{-9}$} & R1 & \multicolumn{1}{c|}{60} & \multicolumn{1}{c|}{3.82} & 8 \\ \cline{4-7} 
 &  &  & R2 & \multicolumn{1}{c|}{140} & \multicolumn{1}{c|}{8.86} & 18 \\ \hline
\multirow{2}{*}{LF} & \multirow{2}{*}{0.01} & \multirow{2}{*}{4.9 x 10$^{-10}$} & R1 & \multicolumn{1}{c|}{180} & \multicolumn{1}{c|}{1.33} & 28 \\ \cline{4-7} 
 &  &  & R2 & \multicolumn{1}{c|}{240} & \multicolumn{1}{c|}{1.64} & 32 \\ \hline
\end{tabular}
\end{table}

Vertical drainage experiments were repeated at two constant flow rates, differing by an order of magnitude. These rates were chosen to represent conditions encountered during typical CO$_2$ migrations in the subsurface near the wellbore (high flow, HF) and at the plume leading edge (low flow, LF) \cite{Jackson2020}. The sample was initially brine saturated, before injection of N\textsubscript{2} from the base at constant flow rate. Constant pressure was maintained at the outlet. The differential pressure response, measuring the difference in pressure from the inlet to the outlet of the rock core, was used to determine when the injected phase had reached the core, triggering continuous imaging until steady state was achieved. Rates and timings of experimental repeats are summarised in Table \ref{tab:SyncExp_timings}. The experiments fall within a capillary dominated regime \cite{Jackson2020REV} and exceed the critical threshold for intermittent fluid connectivity \cite{Reynolds2017}. Full details of the imaging protocol and segmentation workflow are provided in the Supporting Information. Reconstructed X-Ray CT images were registered, filtered and segmented, with N\textsubscript{2} saturations calculated from the segmented images as the fractional pore volume occupied by N\textsubscript{2}.  

\section{Results}

\subsection{Macroscopic saturation distribution}

\begin{figure}[ht!]
\centering
\subfloat[]{\includegraphics[width=0.48\textwidth]{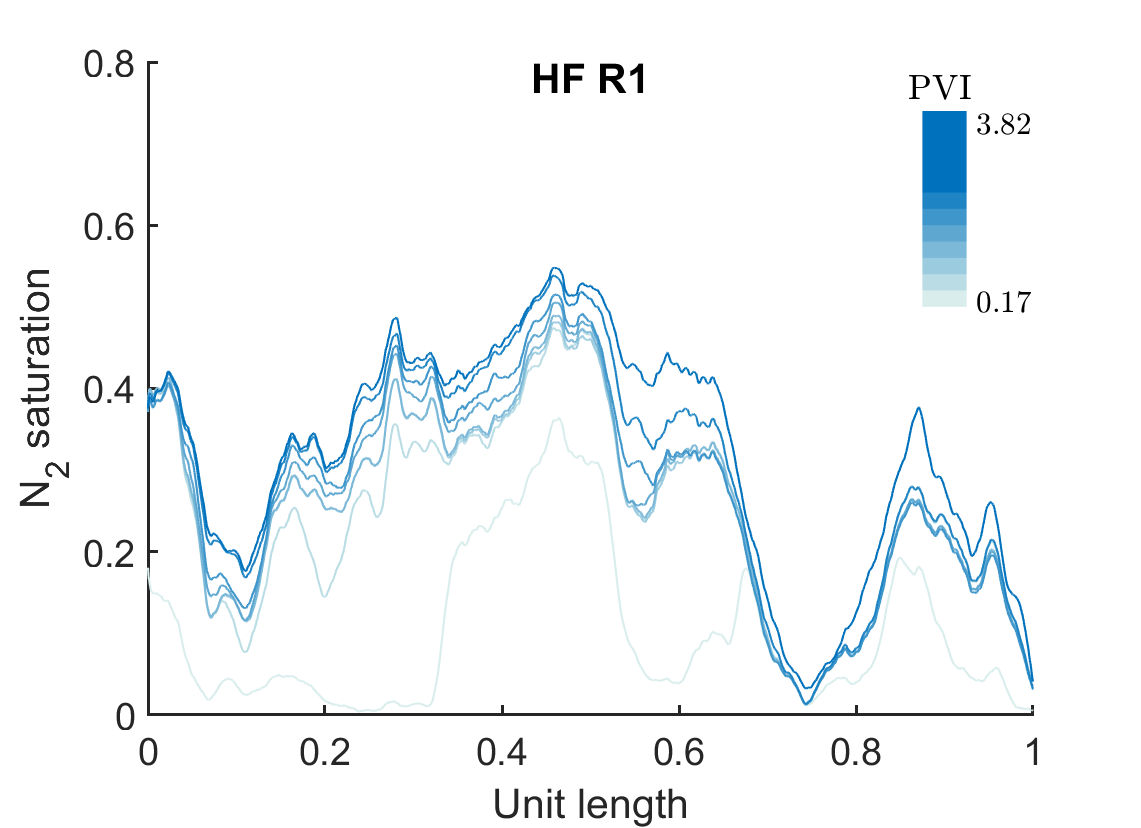}}
\subfloat[]{\includegraphics[width=0.48\textwidth]{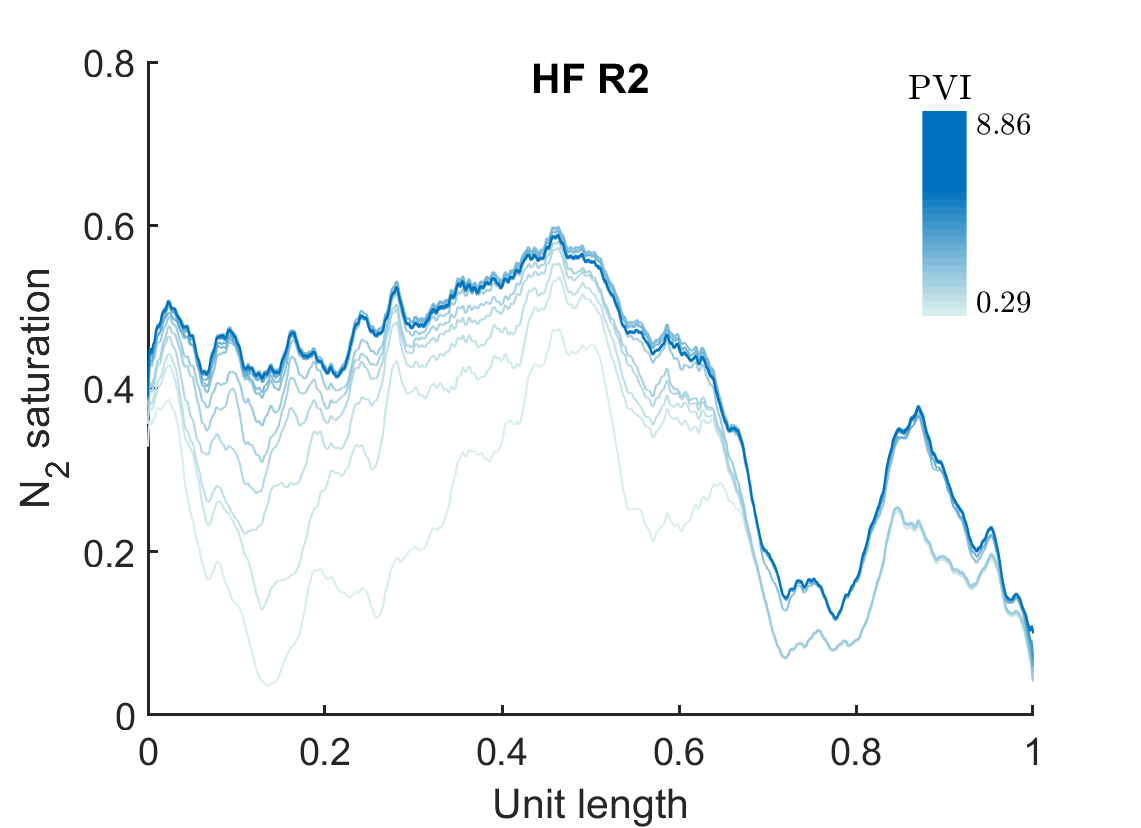}}

\subfloat[]{\includegraphics[width=0.48\textwidth]{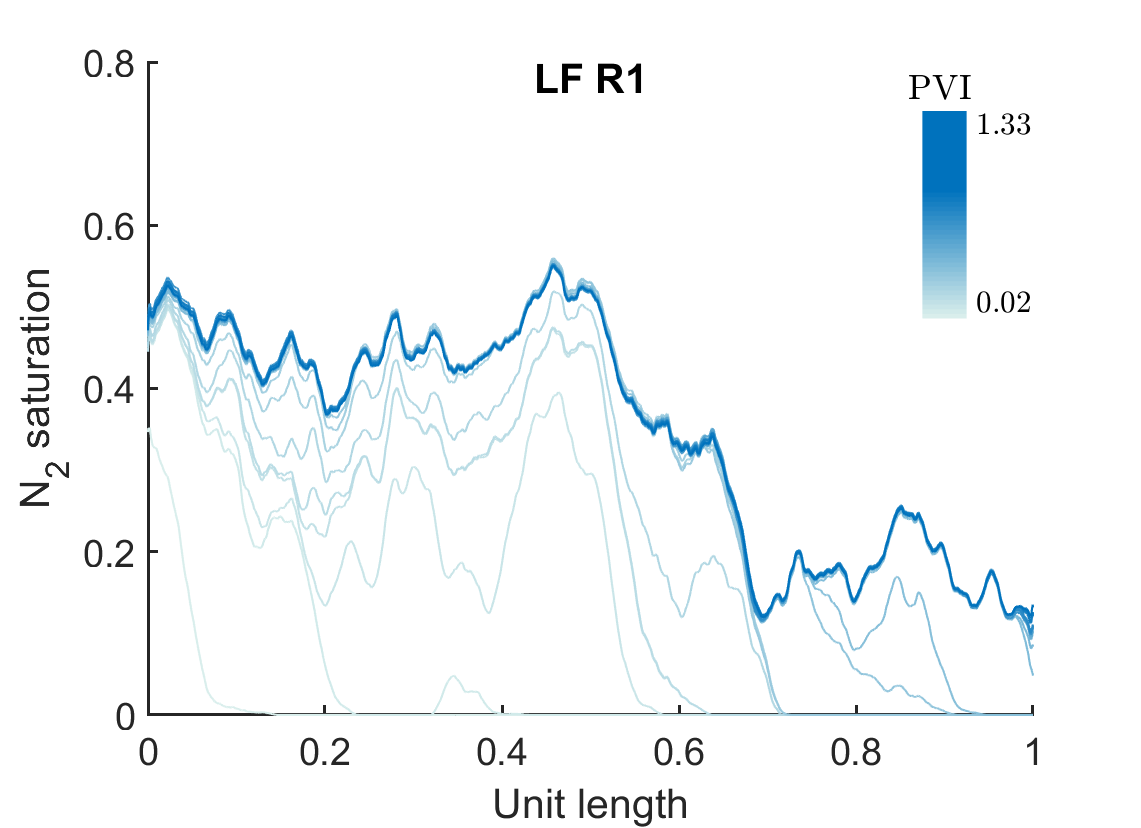}}
\subfloat[]{\includegraphics[width=0.48\textwidth]{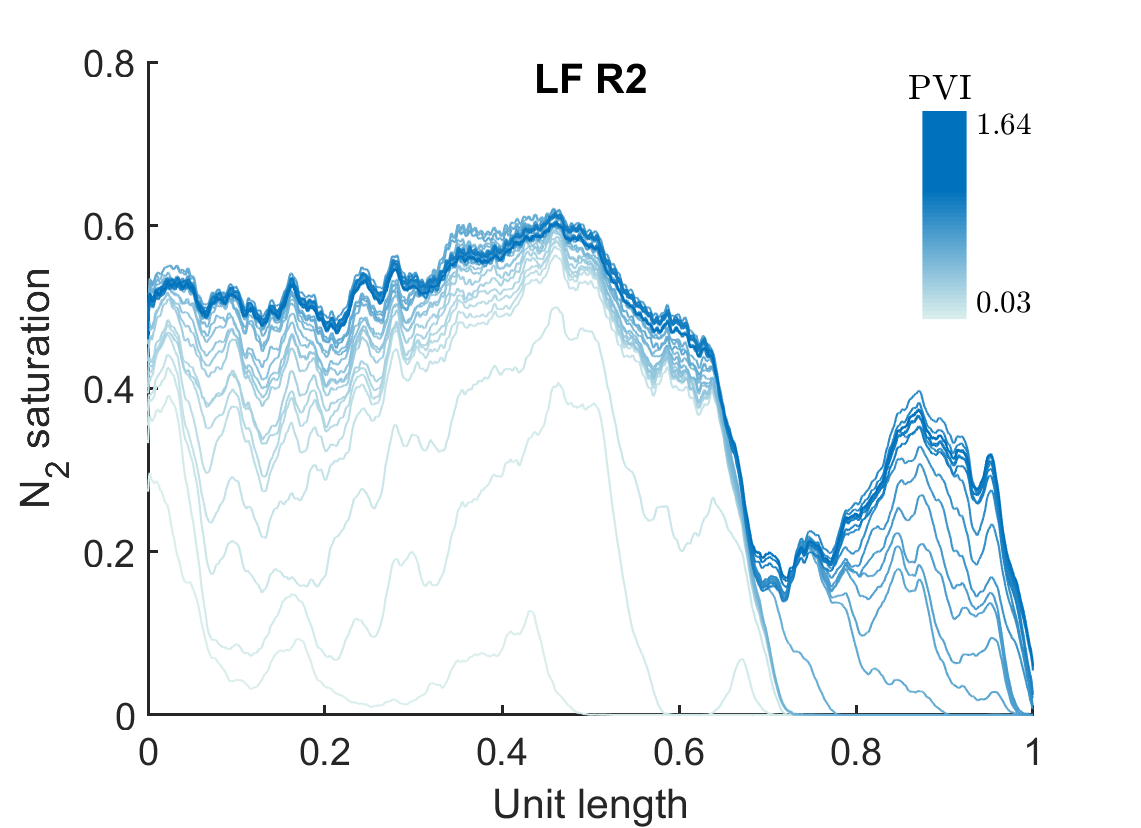}}

\subfloat[]{\includegraphics[width=0.48\textwidth]{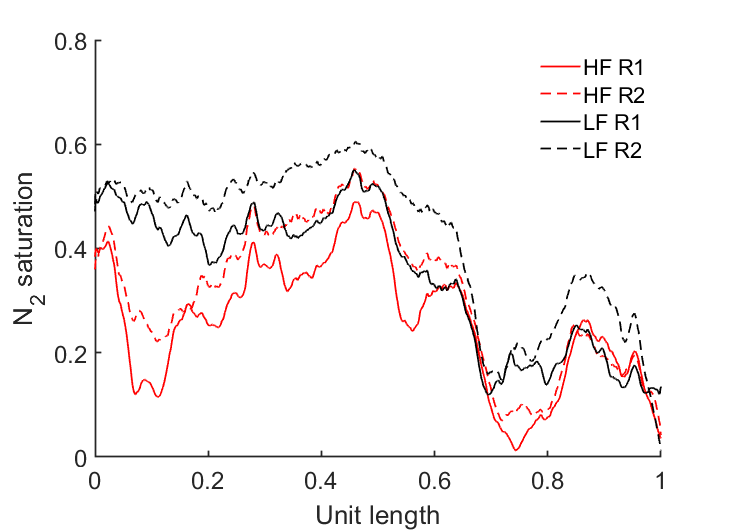}}
\subfloat[]{\includegraphics[width=0.48\textwidth]{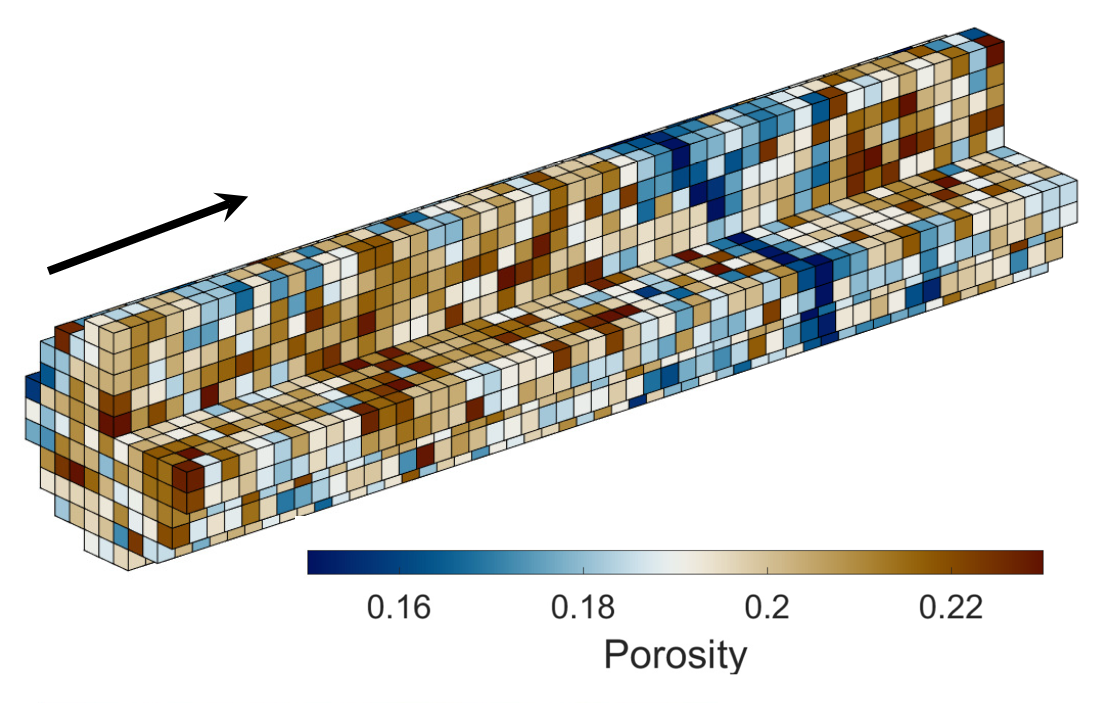}}
\caption{Slice average N\textsubscript{2} saturation along unit core length for high flow rate (10\textsuperscript{-1} ml/min) (a) R1, (b) R2, and low flow rate (10\textsuperscript{-2} ml/min) (c) R1, (d) R2 (with $\sim$ 0.5 and 0.05 pore volumes between scans at high and low flow rate respectively). Each slice has an averaged length of 1mm, chosen to give well-defined REV flow properties.(e) Slice average N\textsubscript{2} saturation along unit core length for repeat experiments after $\sim$ 1.4 pore volumes of N\textsubscript{2} injected. (f) 3D porosity displayed as upscaled $\sim$ 1 mm\textsuperscript{3} voxels. Relative flow direction indicated.}
\label{fig:1Dsat_ANSTOR1}
\end{figure}

The macroscopic N\textsubscript{2} saturation distribution, and its dynamic evolution are displayed in Figure \ref{fig:1Dsat_ANSTOR1}. The sample contained low porosity layers oblique to flow, with a prominent low porosity band two-thirds of the way along the core length, which exerted control over the saturation distribution (Figure \ref{fig:1Dsat_ANSTOR1}f). Across the repeat experiments, the N\textsubscript{2} saturation is elevated upstream of the low porosity band, which acted as a capillary barrier. This is consistent with analytical models and previous observations, which show N\textsubscript{2} will only flow into the lower porosity region when the N\textsubscript{2} saturation built up achieved the capillary entry pressure of the lower porosity region \cite{Dale1997,Dawe2011b}. The elevated N\textsubscript{2} distribution upstream of the heterogeneity persists when the capillary barrier is overcome in order to maintain capillary pressure continuity between the two regions.  

For both repeat experiments, the saturation builds up upstream of the heterogeneity and propagates further upstream at lower flow rates (Figure \ref{fig:1Dsat_ANSTOR1}e). This is consistent with the results observed in numerical models \cite{Harris2021, Dale1997} and those reported in Reynolds et al. (2015) evaluating the impact of capillary number on multiphase flow in heterogeneous rocks \cite{Reynolds2015}. At lower flow rates, when capillary effects dominate, the capillary pressure tends to a constant value with a saturation governed by heterogeneity in the capillary pressure characteristics. At higher flow rates, viscous forces also become important, although the heterogeneity still impacts the saturation distribution. 

At low flow rate, the heterogeneity is observed to impede the breakthrough of the non-wetting phase (Figure \ref{fig:1Dsat_ANSTOR1}c,d), however, the number of pore volumes injected to overcome the heterogeneity varies between low flow rate repeats R1 (0.46 PVI), R2 (0.83 PVI). At high flow rate (Figure \ref{fig:1Dsat_ANSTOR1}a,b), the drainage front advances through the core over a similar time interval to that taken to scan the full core. More pore volumes of N\textsubscript{2} are injected to overcome the heterogeneity at low flow rate, compared to high flow rate, resulting from the additional buildup of N\textsubscript{2} required. These qualitative observations are consistent with numerical simulation results reported in Jackson and Krevor (2020) where capillary heterogeneities act as vertical flow baffles \cite{Jackson2020}, however continuum simulations do not predict a difference in when the N$_2$ overcomes the heterogeneity, for the same petrophysical properties.

Differences in N\textsubscript{2} saturation distribution and its evolution between repeat experiments occur, which would not be seen in continuum simulations. During the first high flow rate experiment (R1, Figure \ref{fig:1Dsat_ANSTOR1}a), post breakthrough, the N\textsubscript{2} saturation continues to increase over many pore volumes injected. In the repeat high flow rate experiment (R2, Figure \ref{fig:1Dsat_ANSTOR1}b), the N\textsubscript{2} saturation increases during breakthrough, before a stable N\textsubscript{2} saturation distribution is observed after $\sim$ 4 PVI. At late times, when high numbers of pore volumes of N\textsubscript{2} have been injected, the fluid redistributes, increasing the N\textsubscript{2} saturation in the downstream region of the core. This results in more N\textsubscript{2} saturated pores within the low porosity region for the R2 experiment, compared to R1. Furthermore, the repeat low flow rate experiment (R2, Figure \ref{fig:1Dsat_ANSTOR1}d) continues to increase in N\textsubscript{2} saturation downstream of the heterogeneity with pore volumes injected, whilst the first experiment (R1, Figure \ref{fig:1Dsat_ANSTOR1}c) remains constant. These results demonstrate that even at very long timescales, the saturation distribution around the heterogeneity can vary when the non-wetting phase continues to be injected. 

\subsection{Pore-scale saturation distribution}

\begin{figure}[H]
\centering
{\includegraphics[width=0.8\textwidth]{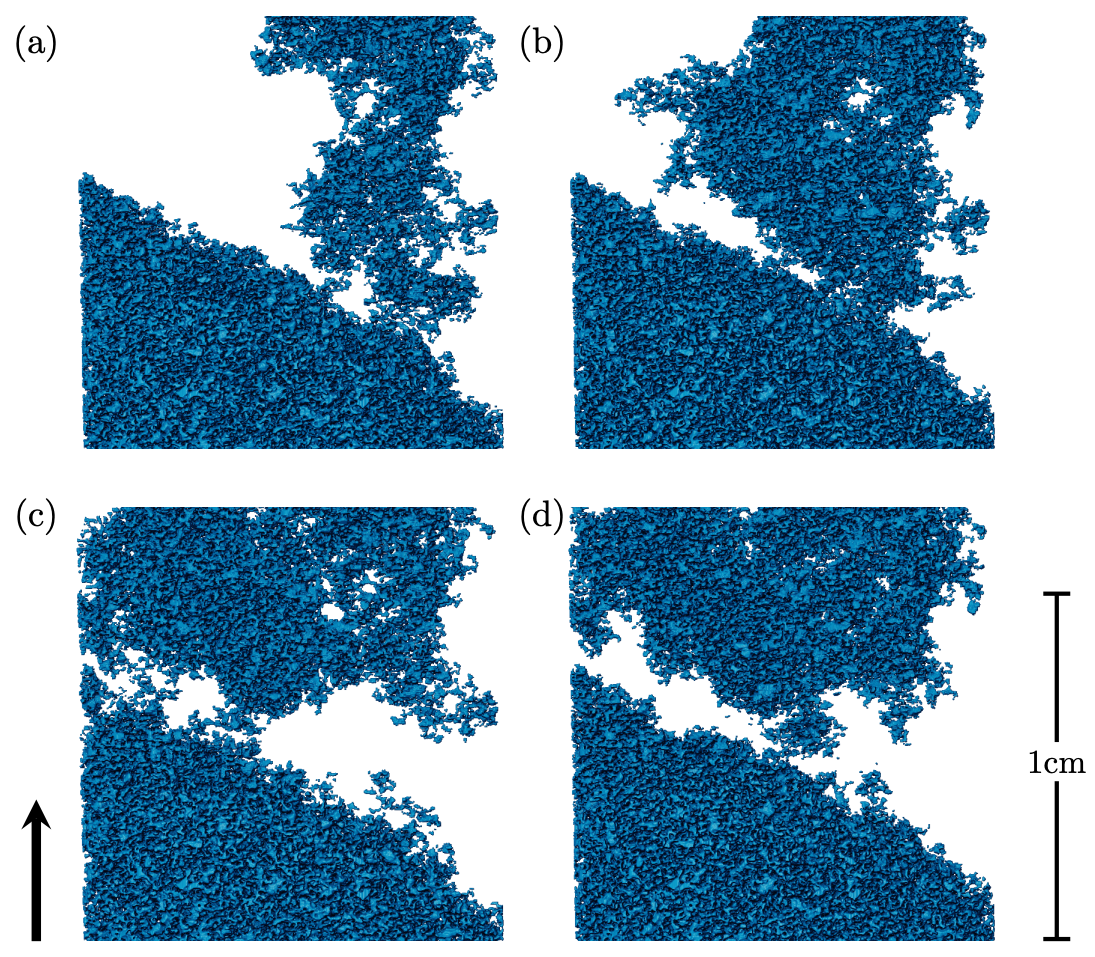}}
\caption{N\textsubscript{2} saturation distribution within the heterogeneity region at the end of drainage for high flow rate (a) R1, (b) R2, and low flow rate (c) R1, (d) R2. The N\textsubscript{2} saturation distribution is displayed as a 3D volume rendering, with the brine and rock phases excluded. Relative flow direction indicated.}
\label{fig:3DDrain_het}
\end{figure}

The underlying connectivity of the pore space influences continuum saturation distributions. The impact of heterogeneity is demonstrated in the 3D volume rendering of the N\textsubscript{2} saturation distribution at the end of drainage (Figure \ref{fig:3DDrain_het}). The flow of N\textsubscript{2} through the core is controlled by just a few pores within the lower porosity layer of the rock, two-thirds of the way along the core length. Different pore-scale flow paths are taken by the N\textsubscript{2} across this layer during repeat experiments. Whilst the low porosity layer has a similar impact on slice-averaged saturation across all the experiments, acting to build up the saturation at an angle dictated by the rock structure, the flow path across the layer varies. The high flow rate experiments (Figure \ref{fig:3DDrain_het}a,b) show pore-scale flow follows a similar path through the layer, but that a different downstream saturation distribution results. The low flow rate experiments (Figure \ref{fig:3DDrain_het}c,d) show a distinct region of the pore space is overcome at the boundary region between the high and low porosity layers, connecting the regions upstream and downstream of the low porosity layer through a different pathway. 

These experiments show that the pore-scale fluid distribution is sensitive to small variations in the conditions below the control of the experiment, even while the macroscopic boundary conditions remain repeatable. The repeat experiments were carried out under the same experimental conditions, on the same core, and therefore could be expected to show identical saturation distributions (See the Supporting Information for validation of the macroscopic volume flux). However, the resulting saturation distributions across the heterogeneity differ. At the centimetre scale (the scale of this sample), several pathways are available to the fluid at each intersection. Often these pathways have a similar energy cost, and thus depend on the smallest of variations in pressure and fluid arrangement to achieve different pore-scale fluid distributions \cite{Bultreys2018}. Small differences in experimental pressure gradients and fluid volume oscillations due to compressibility \cite{Jankov2010} between repeat experiments will exist and it is not possible to remove the influence of these effects from the results. However, the stochastic nature of flow remains the probable cause of these observations.

\subsection{Dynamics of non-wetting phase build up at the capillary barrier}

\begin{figure}[H]
\centering
\subfloat[]{\includegraphics[width=0.55\textwidth]{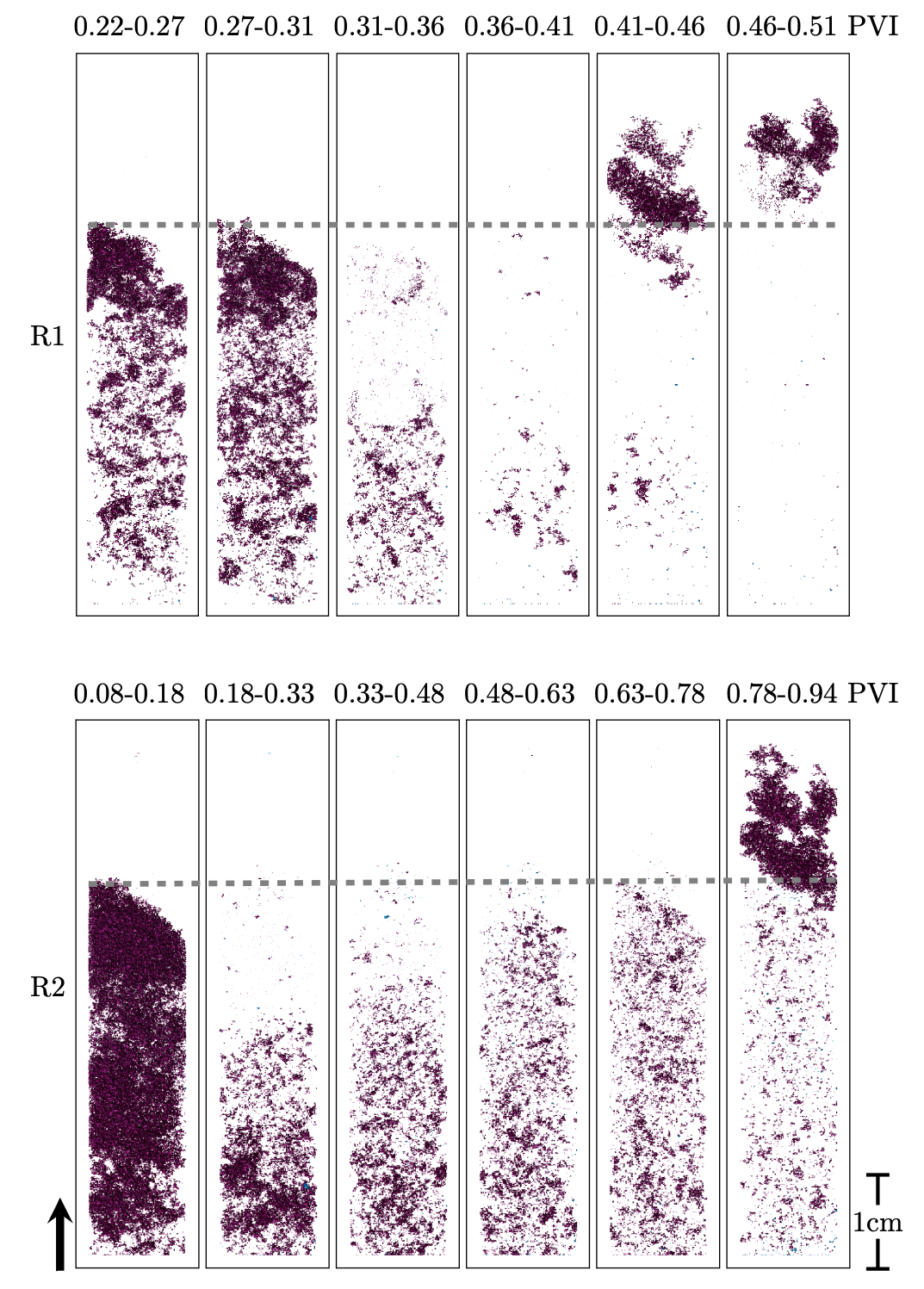}}
\subfloat[]{\includegraphics[width=0.45\textwidth]{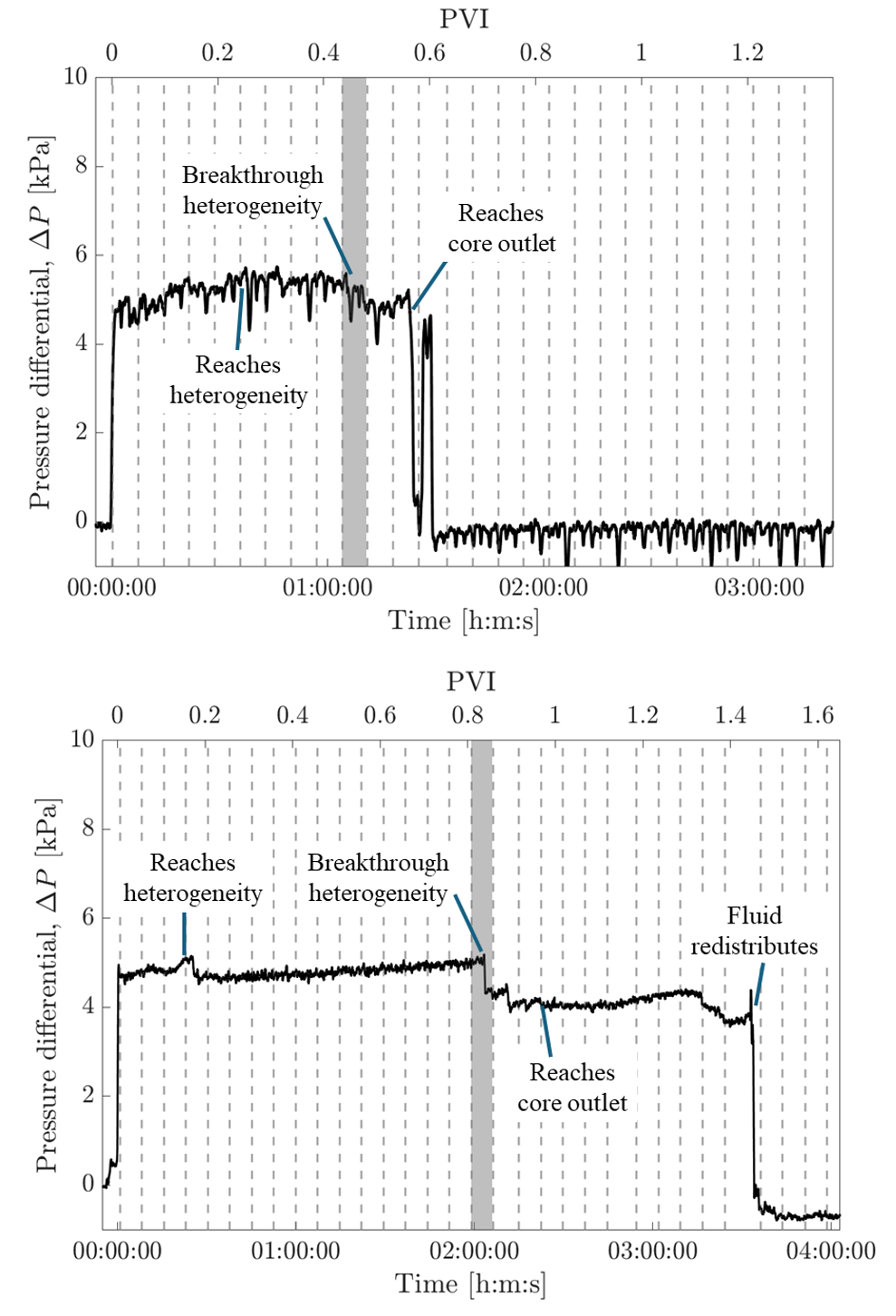}}
\caption{(a) Difference images between scans during low flow rate experiments (R1, top, and R2, bottom). Increases in N\textsubscript{2} are shown in purple, decreases (limited) in blue, with a dashed line marking the heterogeneity. Difference images are calculated at the displayed pore volumes of injected N\textsubscript{2} injected displayed, with timestamps for R2 at 3x longer intervals. (b) Differential pressure response across the core (R1, top, and R2, bottom), averaged over 20 second intervals to reduce noise. Grey shading marks the period during which N\textsubscript{2} breaks through the heterogeneity. Key events are highlighted, with the interval between N\textsubscript{2} reaching the heterogeneity and the core outlet corresponding to the time window shown in the difference images.}
\label{fig:backfilling}
\end{figure}

The experiments demonstrate differences in N$_2$ saturation evolution during drainage, in addition to differences in pressure evolution. The differences in dynamics are most pronounced when comparing the interaction of the N\textsubscript{2} saturation at the low-porosity layer boundary during the low flow rate experiments, due to the greater number of scans captured before breakthrough. An increase in differential pressure is observed alongside the build-up in saturation at the heterogeneity for both repeat low flow rate experiments, indicative of the capillary pressure barrier. When the capillary entry pressure is overcome, a decrease in differential pressure is observed, corresponding with an increased ease of flow (Figure \ref{fig:backfilling}b). However, the details of the pressure response are very different in the two experiments, especially post breakthrough. 

This difference is related to changes in the evolution of the pore-scale saturation distribution, visualised through difference images between consecutive drainage scans (Figure \ref{fig:backfilling}a). This dataset allows us to visualise how N\textsubscript{2} builds up at the boundary between high and low porosity layers at the levels of individual pores in real time. In both repeats, N\textsubscript{2} first built up directly upstream of the low porosity boundary. At later time steps, the N\textsubscript{2} increased further upstream in the core before the boundary was breached. The filling occurs in a complex manner due to the 3D tortuous nature of the pore space. There are fewer changes in the upstream region in the R1 experiments after breakthrough. In contrast,  during the R2 experiment, the N\textsubscript{2} saturation distribution continues to evolve over the entire region through the experiment.

A different response is observed between the repeat experiments, both in the number of pore volumes injected for which the N\textsubscript{2} is baffled and the distribution of N\textsubscript{2} built up (Figure \ref{fig:backfilling}a). During the low flow rate R1 experiment, the N\textsubscript{2} reached the capillary barrier within 0.31 pore volumes injected. An additional 0.15 pore volumes of N\textsubscript{2} was injected whilst the saturation and pressure built up, before the capillary barrier was overcome. For the low flow rate R2 experiment, the N\textsubscript{2} reached the capillary barrier, within 0.18 pore volumes injected. The flow was baffled for 0.65 pore volumes injected before the N\textsubscript{2} establishes a flow path across the low porosity layer. During the R2 experiment, the N\textsubscript{2} backfilling begins at the core inlet, moving through the core as a second wave. Only when this additional N\textsubscript{2} reaches the capillary barrier is it breached. 

The different volumes of N\textsubscript{2} injected to overcome the capillary entry pressure (0.15 PV vs. 0.65 PV, so $\sim$ 4 times longer for R2) are impacted by the N\textsubscript{2} distribution within the core when the low porosity layer is reached. A higher number of pore volumes of N\textsubscript{2} are injected in the R1 experiment before this is reached. The N\textsubscript{2} distribution depends on the relative forces at play, alongside the accessibility and availability of the pore space. These experiments emphasize how stochastic flow across the layer impacts migration speeds through the core. Flow baffling by capillary heterogeneities increases complexity in predicting migration speeds in heterogeneous systems.

% R1: S7 (0.31PV), S10 (0.46 PV), diff 0.15
% R2: S4 (0.18PV), S17 (0.83 PV), diff 0.65

\subsection{Breakthrough the capillary barrier}

The path taken by N\textsubscript{2} across the low-porosity layer reveals how capillary baffles are overcome. Breakthrough occurs once the capillary entry pressure of the controlling throat is exceeded, allowing invasion of the downstream region of the core. To link macroscopic saturation distribution with underlying pore-scale processes, the throats governing flow across the capillary barrier were identified for the low-flow-rate experiments. Notably, N\textsubscript{2} entered the low-porosity region through different pores in the two repeats. 

In the R1 low-flow-rate experiment, breakthrough occurs through a single controlling throat with a diameter measured to be $\sim 30 \pm 10$ $\mu$m. Using the Young-Laplace formula $P_{c} = \frac{2 \gamma cos\theta}{r}$, with interfacial tension $\gamma = 6.98\times10^{-2}$ N/m and $cos\theta = 1$ for a water wet system \cite{Yan2001} the corresponding capillary entry pressure is estimated as $P_{c,e} = 9 \pm 3$ kPa. This value is twice the capillary pressure estimated from the mean, volume-weighted throat radius of the sample (29 $\mu$m, giving 4.8 kPa). The measured differential pressure drop across the system during breakthrough of the low porosity region is 5.5 kPa, underlining the dominant role of smaller throats in controlling flow. In the low flow rate R2 experiment, N$_2$ breaches the capillary pressure barrier through a different part of the pore space. In this case, it is more difficult to isolate the controlling throat because of the large number of pores imaged and the prevalence of intermediate greyscale values caused by fluid motion during the scan. 

Although capillary heterogeneity trapping at the core and continuum scales can enhance the security of underground CO\textsubscript{2} storage \cite{Krevor2011,Krevor2015}, these results highlight its dependence on pore-scale structure and processes. Unlike continuum-scale numerical models, where the capillary barrier is represented as a sharp interface, the real barrier consists of pores with a distribution of sizes and capillary entry pressures. Its influence on fluid migration may be governed by a small subset of pores, as these experiments show, determined by the path of least resistance that is both available and accessible to the invading phase. This suggests that when using digital rock models to represent capillary heterogeneity at the continuum scale, care must be taken to capture the properties of the minimum barrier to flow.

\begin{figure}[h]
\centering
\subfloat[]{\includegraphics[width=0.4\textwidth]{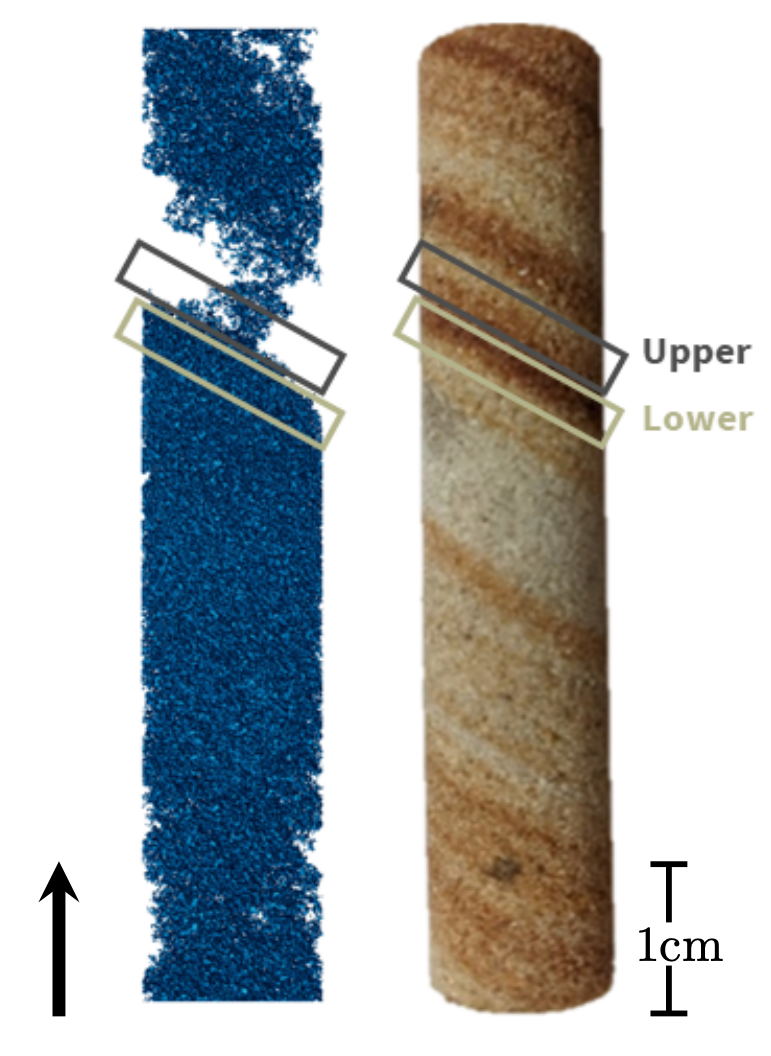}}
\subfloat[]{\includegraphics[width=0.48\textwidth]{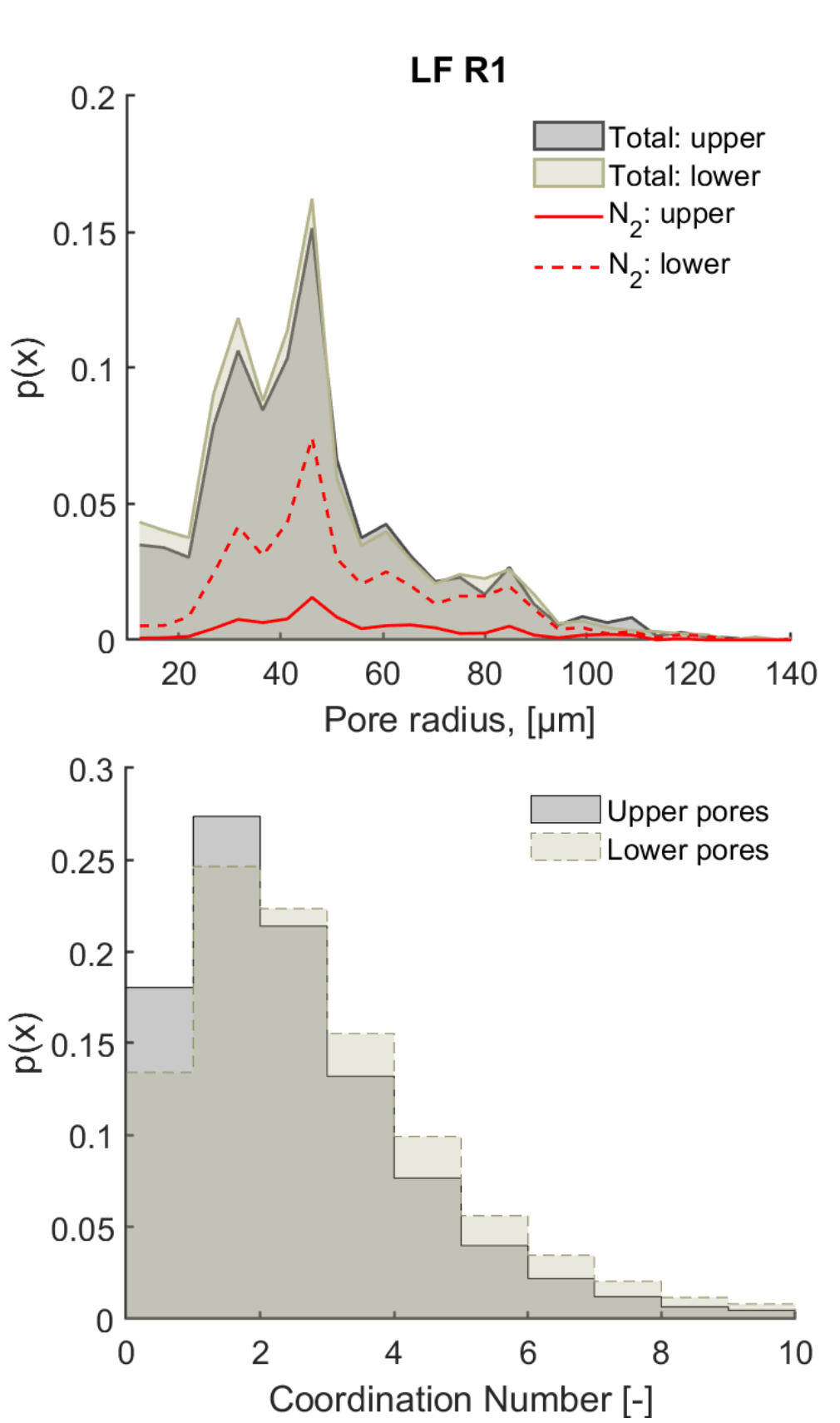}}
\caption{(a) Upper and lower boundary regions on a 3D volume rending of low flow rate R1 N\textsubscript{2} saturation (end of drainage) and an image of the core. (b) top -  Pore size distribution (bodies and throats) within the upper and lower boundary region defined for the R1 experiment, overlaid by N\textsubscript{2} occupied pores at the end of low flow rate drainage in each region. The probability is relative to the total number of pores and throats within the lower region. Bottom - Coordination number distribution of the upper and lower region pores.}
\label{fig:HetUpperLower}
\end{figure}

To investigate pore-scale controls on fluid distribution, pore networks were extracted from two regions adjacent to the capillary barrier: one upstream and one downstream, identified by the clear layer of N\textsubscript{2} accumulation (Figure \ref{fig:HetUpperLower}a). In the R1 low-flow-rate experiment, we compare pore size distributions and fluid occupancy across these boundary layers. Figure \ref{fig:HetUpperLower}b shows that the total pore size distributions (bodies and throats) are broadly similar, with only a $\sim$10\% difference in the total number of pores and throats. Averaged over $\sim$20,000 extracted pores, both regions therefore exhibit comparable global pore size distributions, with nearly identical mean pore radii (37.8 $\mu$m vs. 37.9 $\mu$m). Despite this similarity, Figure \ref{fig:HetUpperLower} demonstrates markedly different saturation behaviour, indicating that regions with comparable bulk properties can exhibit distinct flow behaviour.

Subtle differences in pore structure help explain differences in flow. The upper region has a lower porosity ($\phi$ = 0.168) than the lower region ($\phi$ = 0.180), defining the heterogeneity. Pore and throat sizes are also reduced: the maximum pore size is 125 $\mu$m in the upper region versus 175 $\mu$m in the lower, and the maximum throat size is 72 $\mu$m versus 92 $\mu$m. In addition, pore connectivity is reduced. The upper region has a throat-to-pore ratio of 1.0 and an average coordination number of 2.1, compared with 1.2 and 2.5, respectively, for the lower region. Thus, while pore size distributions appear similar, the absence of larger, well-connected pores in the upper region creates the observed capillary barrier to N\textsubscript{2}.

These results show that even subtle variations in pore structure can strongly influence fluid distribution and trapping. The impact of a capillary barrier depends more on pore arrangement and connectivity than on global statistics such as pore-size distribution. The observed reduction in both porosity and connectivity at the heterogeneity limits the pathways available for N\textsubscript{2} migration. This supports the interpretation that capillary heterogeneity trapping is primarily an interfacial condition rather than a volumetric one \cite{Matthai2023}.

\section{Discussion}

\subsection{Unstable nature of fluid migration}

% P1 - flow is chaotic 
%P2 - Chaos is direct result of heterogeneity

Pore-scale flow is strongly influenced by macroscopic heterogeneities, which control saturation distributions at the macro-scale. Repeated experiments exhibit different macroscopic saturation distributions and dynamics, highlighting the role of heterogeneity in generating unstable fluid distributions. At the pore-scale, many flow paths are similar from an energetic perspective. Small perturbations—linked to gas compressibility and amplified by its low viscosity relative to water—can therefore cause different interactions with the heterogeneity, impacting macroscopic saturation distributions. Reduced connectivity and porosity at boundaries between regions are identified as the key parameters controlling flow, even when pore-size distributions appear similar.

We observe that in heterogeneous systems, fluid distributions are inherently unstable. In homogeneous rocks, repeatability has been examined by Bultreys et al. (2018), who showed that while pore-scale details varied, global saturation properties were reproducible. Our results go further, demonstrating that heterogeneity drives differences in global properties including fluid distribution and dynamics. This finding supports the non-uniqueness reported in other heterogeneous systems \cite{Fernø2023}. 

\subsection{Implications at the field-scale}

%P1 - Model stchastically
Beyond advancing pore-scale understanding, these results have direct consequences for dynamic modelling of CO\textsubscript{2} storage, among other environmental and engineering processes. They show that macroscopic flow patterns may be sensitive to small perturbations or local flow conditions at heterogeneities, thus, their representation should consider the use of stochastic models rather than deterministic ones. An uncertainty approach should be developed taking into account the uncertainties resulting from these small scale heterogeneities as well as uncertainties in the distribution of larger scale heterogeneities. Capturing the full range of flow behaviours may require multiple realisations with varying boundary conditions. Such stochastic models are particularly relevant for predictions of geological storage in field sites such as Sleipner, where deterministic simulations struggle to predict CO\textsubscript{2} distribution and breakthrough times due to interactions with imperfect sealing layers \cite{Chadwick2009}.   

Different macroscopic saturations between repeat experiments result from a graduated pore-size variation at the boundary, producing a distribution of capillary entry pressures. This contrasts with the sharp-interface assumption commonly adopted in continuum-scale models. The different flow paths impact the amount of fluid that must be injected to pass into the lower porosity region and move through the core (a factor of $\sim$ 4 difference between repeats). This, combined with the adverse viscosity ratio between gas and brine, has implications when predicting migration speeds in the field. It may be particularly important in subsurface applications that rely upon capillary heterogeneity for migration-assisted trapping, sometimes in lieu of structural trapping such as in composite confining systems \cite{Bump2023,NI2024104125}.

% P3 - flow through small number of pores has implications for Pc imb curve
Fluid distributions through a lower porosity region are repeatedly characterised by percolation through only a few critical pores and flow pathways. This has direct implications for the continuum description of heterogeneity as snap-off in these pores links directly to the magnitude of the snap-off pressure in a continuum capillary pressure imbibition curve. Further experiments are needed to investigate trapping dynamics during imbibition.

\subsection{Overcoming imaging challenges}

%P1 - Resolution constraints impact intermittency measurements
%P2 - Large samples encouraged, but challenges

Processes spanning multiple spatial and temporal scales are ubiquitous across both natural and engineered systems. By demonstrating the role of heterogeneity on fluid distribution and dynamics at the centimeter-scale, our study underscores the need for representative sample sizes to capture average behaviour. However, resolving individual pores and simultaneously capturing dynamics across the macro-scale imposes resolution constraints. Capturing intermittency over this timescale is particularly challenging, therefore understanding how it interacts with macroscopic heterogeneities, especially when flow is controlled by only a few pores, remains a key research question.

Next-generation synchrotrons, with higher brilliance and coherence, offer new opportunities for resolving such dynamics \cite{Chapman2023}. Yet, they also present challenges associated with the large datasets generated and the complexity of analysing evolving, large-scale pore networks. Progress in field-of-view, synchrotron photon flux, and big-data image analysis will be critical for improving dynamic measurements of macro-scale heterogeneity \cite{Goethals2025,wang2021}.

In this study, we demonstrate how a synchrotron beamline - previously unused for this type of experiment - can capture multiscale flow processes in real time. Our combined experimental and analysis approach demonstrates new opportunities for investigating multiscale dynamics in complex, opaque porous media and highlights the potential of leveraging diverse beamlines to address cross-disciplinary research challenges.

\section*{Open Research Section}
The X-Ray micro-CT data, saturation, pressure and pore-network data used in the study in the study are available at https://zenodo.org/ via [DOI to appear on publication] with a Creative Commons Attribution 4.0 International license \cite{Harris2025}.

\section*{Conflict of Interest declaration}
The authors declare there are no conflicts of interest for this manuscript.

\acknowledgments
We gratefully acknowledge sponsorship from the EPSRC (Grant Number EP/R513052/1) and BP. 

\bibliography{references.bib}

@article{Jackson2020,
author = {Jackson, Samuel J. and Krevor, Samuel},
doi = {10.1029/2020GL088616},
journal = {Geophysical Research Letters},
title = {{Small‐Scale Capillary Heterogeneity Linked to Rapid Plume Migration During CO2 Storage}},
volume = {47},
number = {18},
pages = {e2020GL088616},
year = {2020}
}

@Article{Moura2020,
  author       = {Moura, Marcel and Måløy, Knut Jørgen and Flekkøy, Eirik Grude and Toussaint, Renaud},
  date         = {2020-01},
  journal = {Frontiers in Physics},
  title        = {Intermittent Dynamics of Slow Drainage Experiments in Porous Media: Characterization Under Different Boundary Conditions},
  doi          = {10.3389/fphy.2019.00217},
  issn         = {2296-424X},
  volume       = {7},
  publisher    = {Frontiers Media SA},
}

@article{Zhao2016,
author = {Benzhong Zhao  and Christopher W. MacMinn  and Ruben Juanes },
title = {Wettability control on multiphase flow in patterned microfluidics},
journal = {Proceedings of the National Academy of Sciences},
volume = {113},
number = {37},
pages = {10251-10256},
year = {2016},
doi = {10.1073/pnas.1603387113}}

@Article{Chapman2023,
  author       = {Chapman, Henry N.},
  year         = {2023},
  journal = {IUCrJ},
  title        = {Fourth-generation light sources},
  doi          = {10.1107/s2052252523003585},
  issn         = {2052-2525},
  number       = {3},
  pages        = {246--247},
  volume       = {10},
  publisher    = {International Union of Crystallography (IUCr)},
}

@ARTICLE{Goethals2025,
  author={Goethals, Wannes and Bultreys, Tom and Berg, Steffen and Boone, Matthieu N. and Aelterman, Jan},
  journal={IEEE Transactions on Computational Imaging}, 
  title={DYRECT Computed Tomography: DYnamic Reconstruction of Events on a Continuous Timescale}, 
  year={2025},
  volume={11},
  number={},
  pages={638-649},
  doi={10.1109/TCI.2025.3566241}}

@ARTICLE{Harris2025,
author = {Harris, Catrin and  Muggeridge, Ann H and Krevor, Samuel and Camilerri, Michael and Jackson, Samuel J.},
  journal={Zenodo}, 
  title={Dynamic X-Ray micro-CT images of gas-brine flow in Bentheimer Sandstone (Version 1) [Data set]}, 
  year={2025}}

@Article{Jankov2010,
  author       = {Jankov, M. and Løvoll, G. and Knudsen, H. A. and Måløy, K. J. and Planet, R. and Toussaint, R. and Flekkøy, E. G.},
  year         = {2010},
  journal = {Transport in Porous Media},
  title        = {Effects of Pressure Oscillations on Drainage in an Elastic Porous Medium},
  doi          = {10.1007/s11242-009-9521-z},
  number       = {3},
  pages        = {569--585},
  volume       = {84},
  publisher    = {Springer Science and Business Media LLC},
}

@article{maloy1992,
  title = {Dynamics of slow drainage in porous media},
  author = {M\aa{}l\o{}y, Knut J\o{}rgen and Furuberg, Liv and Feder, Jens and J\o{}ssang, Torstein},
  journal = {Phys. Rev. Lett.},
  volume = {68},
  issue = {14},
  pages = {2161--2164},
  numpages = {0},
  year = {1992},
  month = {Apr},
  publisher = {American Physical Society},
  doi = {10.1103/PhysRevLett.68.2161},
}

@article{chadwick2015,
author = {Chadwick, R. Andrew and Noy, David J.},
title = {Underground {CO2} storage: demonstrating regulatory conformance by convergence of history-matched modeled and observed {CO2} plume behavior using Sleipner time-lapse seismics},
journal = {Greenhouse Gases: Science and Technology},
volume = {5},
number = {3},
pages = {305-322},
doi = {https://doi.org/10.1002/ghg.1488},
year = {2015}
}

@article{Dale1997,
author = {Dale, Magnar and Ekrann, Steinar and Mykkeltveit, Johannes and Virnovsky, George},
doi = {10.1023/A:1006536021302},
journal = {Transport in Porous Media},
title = {{Effective Relative Permeabilities and Capillary Pressure for One-Dimensional Heterogeneous Media}},
volume = {26},
number = {3},
pages = {229--260},
year = {1997}
}

@article{Saadatpoor2010,
author = {Saadatpoor, Ehsan and Bryant, Steven L. and Sepehrnoori, Kamy},
doi = {10.1007/s11242-009-9446-6},
journal = {Transport in Porous Media},
title = {{New Trapping Mechanism in Carbon Sequestration}},
volume = {82},
number = {1},
pages = {3--17},
year = {2010}
}

@article{Krevor2015,
author = {Krevor, Samuel and Blunt, Martin J. and Benson, Sally M. and Pentland, Christopher H. and Reynolds, Catriona and Al-Menhali, Ali and Niu, Ben},
doi = {10.1016/j.ijggc.2015.04.006},
journal = {International Journal of Greenhouse Gas Control},
title = {{Capillary trapping for geologic carbon dioxide storage - From pore scale physics to field scale implications}},
volume = {40},
pages = {221--237},
year = {2015}
}

@phdthesis{Regnier2019,
author = {Regnier, Geraldine},
school = {Imperial College London},
title = {{Modelling of CO\textsubscript{2} plume dynamics at the lab-scale using an invasion percolation algorithm}},
year = {2019}
}

@article{Kortekaas1985,
author = {Kortekaas, T. F. M.},
doi = {10.2118/12112-PA},
journal = {Society of Petroleum Engineers Journal},
title = {{Water/Oil Displacement Characteristics in Crossbedded Reservoir Zones}},
volume = {25},
number = {6},
pages = {917--926},
year = {1985}
}

@article{Bech2018,
author = {Bech, Niels and Frykman, Peter},
doi = {10.1016/j.ijggc.2018.06.018},
journal = {International Journal of Greenhouse Gas Control},
title = {{Trapping of buoyancy-driven CO\textsubscript{2} during imbibition}},
volume = {78},
pages = {48--61},
year = {2018}
}

@article{Krevor2011, 
author = {Krevor, Samuel and Pini, Ronny and Li, Boxiao and Benson, Sally M.},
doi = {10.1029/2011GL048239},
journal = {Geophysical Research Letters},
title = {{Capillary heterogeneity trapping of CO\textsubscript{2} in a sandstone rock at reservoir conditions}},
volume = {38},
number = {15},
pages = {L15401},
year = {2011}
}

@article{Jackson2018, 
author = {Jackson, Samuel J. and Agada, Simeon and Reynolds, Catriona A. and Krevor, Samuel},
doi = {10.1029/2017WR022282},
journal = {Water Resources Research},
pages = {3139--3161},
title = {{Characterizing Drainage Multiphase Flow in Heterogeneous Sandstones}},
volume = {54},
number = {4},
year = {2018}
}

@article{Jackson2020REV, 
author = {Jackson, Samuel J. and Lin, Q. and Krevor, S.},
doi = {10.1029/2019wr026396},
journal = {Water Resources Research},
title = {{Representative Elementary Volumes, Hysteresis and Heterogeneity in Multiphase Flow From the Pore to Continuum Scale}},
volume = {56},
number = {6},
pages = {e2019WR026396},
year = {2020}
}

@article{Niu2015,
author = {Niu, Ben and Al‐Menhali, Ali and Krevor, Samuel C.},
doi = {10.1002/2014WR016441},
journal = {Water Resources Research},
pages = {2009--2029},
title = {{The impact of reservoir conditions on the residual trapping of carbon dioxide in Berea sandstone}},
volume = {51},
number = {4},
year = {2015}
}

@article{Li2015,
author = {Li, Boxiao and Benson, Sally M.},
doi = {10.1016/j.advwatres.2015.07.010},
journal = {Advances in Water Resources},
title = {{Influence of small-scale heterogeneity on upward CO\textsubscript{2} plume migration in storage aquifers}},
volume = {83},
pages = {389--404},
year = {2015}
}

@article{Reynolds2015,
author = {Reynolds, C. A. and Krevor, S.},
doi = {10.1002/2015WR018046},
journal = {Water Resources Research},
title = {{Characterizing flow behavior for gas injection: Relative permeability of CO\textsubscript{2}-brine and N\textsubscript{2}-water in heterogeneous rocks}},
volume = {51},
number = {12},
pages = {9464--9489},
year = {2015}
}

@article{Chadwick2009,
author = {Chadwick, R. A. and Noy, D. and Arts, R. and Eiken, O.},
doi = {10.1016/j.egypro.2009.01.274},
journal = {Energy Procedia},
publisher = {Elsevier},
title = {{Latest time-lapse seismic data from Sleipner yield new insights into CO\textsubscript{2} plume development}},
volume = {1},
number = {1},
pages = {2103--2110},
year = {2009}
}

@article{Ringrose1993,
author = {Ringrose, P S and Sorbie, K S and Corbett, P W M and Jensen, J L},
title = {{Immiscible flow behaviour in laminated and cross-bedded sandstones}},
doi = {10.1016/0920-4105(93)90071-L},
journal = {Journal of Petroleum Science and Engineering},
volume = {9},
number = {2},
pages = {103--124},
year = {1993}
}

@article{Zahasky2020a,
author = {Zahasky, Christopher and Jackson, Samuel J. and Lin, Qingyang and Krevor, Samuel},
doi = {10.1029/2019WR026708},
journal = {Water Resources Research},
title = {{Pore Network Model Predictions of Darcy-Scale Multiphase Flow Heterogeneity Validated by Experiments}},
volume = {56},
number = {6},
pages = {1--16},
year = {2020}
}

@article{Spurin2020,
title = {Real-{Time} {Imaging} {Reveals} {Distinct} {Pore}-{Scale} {Dynamics} {During} {Transient} and {Equilibrium} {Subsurface} {Multiphase} {Flow}},
volume = {56},
doi = {10.1029/2020WR028287},
number = {12},
journal = {Water Resources Research},
author = {Spurin, Catherine and Bultreys, Tom and Rücker, Maja and Garfi, Gaetano and Schlepütz, Christian M. and Novak, Vladimir and Berg, Steffen and Blunt, Martin J. and Krevor, Samuel},
year = {2020},
pages = {e2020WR028287},
}

@article{Singh2017,
author = {Singh, Kamaljit and Menke, Hannah and Andrew, Matthew and Lin, Qingyang and Rau, Christoph and Blunt, Martin J. and Bijeljic, Branko},
doi = {10.1038/s41598-017-05204-4},
journal = {Scientific Reports},
title = {{Dynamics of snap-off and pore-filling events during two-phase fluid flow in permeable media}},
volume = {7},
number = {1},
pages = {5192},
year = {2017}
}

@article{Reynolds2017,
author = {Reynolds, Catriona A. and Menke, Hannah and Andrew, Matthew and Blunt, Martin J. and Krevor, Samuel},
doi = {10.1073/pnas.1702834114},
journal = {Proceedings of the National Academy of Sciences},
title = {{Dynamic fluid connectivity during steady-state multiphase flow in a sandstone}},
volume = {114},
number = {31},
pages = {8187--8192},
year = {2017}
}

@article{Harris2021,
author = {Harris, Catrin and Jackson, Samuel J. and Benham, Graham P. and Krevor, Samuel and Muggeridge, Ann H.},
doi = {10.1016/j.ijggc.2021.103511},
journal = {International Journal of Greenhouse Gas Control},
title = {{The impact of heterogeneity on the capillary trapping of CO\textsubscript{2} in the Captain Sandstone}},
volume = {112},
pages = {103511},
year = {2021}
}

@article{Rucker2015,
author = {R{\"{u}}cker, M. and Berg, S. and Armstrong, R. T. and Georgiadis, A. and Ott, H. and Schwing, A. and Neiteler, R. and Brussee, N. and Makurat, A. and Leu, L. and Wolf, M. and Khan, F. and Enzmann, F. and Kersten, M.},
doi = {10.1002/2015GL064007},
journal = {Geophysical Research Letters},
title = {{From connected pathway flow to ganglion dynamics}},
volume = {42},
number = {10},
pages = {3888--3894},
year = {2015}
}

@article{Peksa2015,
author = {Peksa, Anna E. and Wolf, Karl-Heinz A.A. and Zitha, Pacelli L.J.},
doi = {10.1016/j.marpetgeo.2015.06.001},
journal = {Marine and Petroleum Geology},
pages = {701--719},
publisher = {Elsevier Ltd},
title = {{Bentheimer sandstone revisited for experimental purposes}},
volume = {67},
year = {2015}
}

@article{Yan2001,
author = {Yan, W. and Zhao, G. Y. and Chen, G. J. and Guo, T. M.},
doi = {10.1021/je0101505},
journal = {Journal of Chemical and Engineering Data},
title = {{Interfacial Tension of (Methane + Nitrogen) + Water and (Carbon Dioxide + Nitrogen) + Water Systems}},
volume = {46},
number = {6},
pages = {1544--1548},
year = {2001}
}

@article{Seyyedi2022,
author = {Seyyedi, Mojtaba and Clennell, Michael Benedict and Jackson, Samuel J.},
doi = {10.1016/j.advwatres.2022.104216},
journal = {Advances in Water Resources},
pages = {{104216}},
title = {{Time-lapse imaging of flow instability and rock heterogeneity impacts on CO\textsubscript{2} plume migration in meter long sandstone cores}},
volume = {164},
year = {2022}
}

@article{Ni2021,
author = {Ni, Hailun and M{\o}yner, Olav and Kurtev, Kuncho D. and Benson, Sally M.},
doi = {10.1016/j.advwatres.2021.103990},
journal = {Advances in Water Resources},
title = {{Quantifying CO\textsubscript{2} capillary heterogeneity trapping through macroscopic percolation simulation}},
volume = {155},
pages = {103990},
year = {2021}
}

@article{Ni2023,
author = {Ni, Hailun and Bakhshian, Sahar and Meckel, T. A.},
doi = {10.1038/s41598-023-29360-y},
journal = {Scientific Reports},
title = {{Effects of grain size and small-scale bedform architecture on CO\textsubscript{2} saturation from buoyancy-driven flow}},
volume = {13},
number = {1},
pages = {{2474}},
year = {2023}
}

@article{Bump2023,
author = {Bump, Alexander P. and Bakhshian, Sahar and Ni, Hailun and Hovorka, Susan D. and Olariu, Marianna I. and Dunlap, Dallas and Hosseini, Seyyed A. and Meckel, Timothy A.},
doi = {10.1016/j.ijggc.2023.103908},
journal = {International Journal of Greenhouse Gas Control},
pages = {{103908}},
title = {{Composite confining systems: Rethinking geologic seals for permanent CO\textsubscript{2} sequestration}},
volume = {126},
year = {2023}
}

@article{Jackson2022,
author = {Jackson, Samuel J. and Niu, Yufu and Manoorkar, Sojwal and Mostaghimi, Peyman and Armstrong, Ryan T.},
doi = {10.1103/PhysRevApplied.17.054046},
journal = {Physical Review Applied},
title = {{Deep Learning of Multiresolution X-Ray Micro-Computed-Tomography Images for Multiscale Modeling}},
volume = {17},
number = {5},
pages = {054046},
year = {2022}
}

@article{Wang2021,
author = {Wang, Ying Da and Shabaninejad, Mehdi and Armstrong, Ryan T. and Mostaghimi, Peyman},
doi = {10.1016/j.asoc.2021.107185},
journal = {Applied Soft Computing},
pages = {107185},
title = {{Deep neural networks for improving physical accuracy of 2D and 3D multi-mineral segmentation of rock micro-CT images}},
volume = {104},
year = {2021}
}

@incollection{Fernø2023,
    author = {Fern{\o}, M. A. and Haugen, M. and Eikehaug, K. and Folkvord, O. and Benali, B. and Both, J. W. and Storvik, E. and Nixon, C. W. and Gawthrope, R. L. and Nordbotten, J. M.},
    title = {{Room-Scale CO{\textsubscript{2}} Injections in a Physical Reservoir Model with Faults}},
    booktitle = {Special Issue: FluidFlower: A Meter-scale Experimental Laboratory for Geological CO{\textsubscript{2}} Storage},
    editor = {Fern{\o}, M. A. and Flemisch, B. and Juanes R. and Nordbotten, J. M.},
    publisher = {Transport in Porous Media},
    year = {2023},
    doi = {10.1007/s11242-023-02013-4}
}

@article{Bultreys2018,
author = {Bultreys, Tom and Lin, Qingyang and Gao, Ying and Raeini, Ali Q. and Alratrout, Ahmed and Bijeljic, Branko and Blunt, Martin J.},
doi = {10.1103/PhysRevE.97.053104},
journal = {Physical Review E},
title = {{Validation of model predictions of pore-scale fluid distributions during two-phase flow}},
volume = {97},
number = {5},
pages = {053104},
year = {2018}
}

@article{Matthai2023,
author = {Matthai, Stephan K. and Tran, Luat K.},
doi = {10.1016/j.advwatres.2023.104430},
journal = {{Advances in Water Resources}},
title = {{Numeric determination of relative permeability of heterogeneous porous media with capillary discontinuities}},
volume = {175},
pages = {104430},
year = {2023}
}

@article{Dawe2011b,
author = {Dawe, Richard A. and Caruana, Albert and Grattoni, Carlos A.},
doi = {10.1007/s11242-010-9642-4},
issn = {01693913},
journal = {Transport in Porous Media},
number = {2},
pages = {601--616},
title = {{Microscale Visual Study of End Effects at Permeability Discontinuities}},
volume = {86},
year = {2011}
}

@article{Trevisan2017a,
author = {Trevisan, Luca and Pini, Ronny and Cihan, Abdullah and Birkholzer, Jens T. and Zhou, Quanlin and Gonz{\'{a}}lez-Nicol{\'{a}}s, Ana and Illangasekare, Tissa H.},
doi = {10.1002/2016WR019749},
journal = {Water Resources Research},
pages = {485--502},
title = {{Imaging and quantification of spreading and trapping of carbon dioxide in saline aquifers using meter-scale laboratory experiments}},
volume = {53},
year = {2017}
}

@article{NI2024104125,
title = {An experimental investigation on the CO2 storage capacity of the composite confining system},
journal = {International Journal of Greenhouse Gas Control},
volume = {134},
pages = {104125},
year = {2024},
issn = {1750-5836},
doi = {https://doi.org/10.1016/j.ijggc.2024.104125},
author = {Hailun Ni and Alexander P. Bump and Sahar Bakhshian},
}

\end{document}